\documentclass[pre,twocolumn,a4paper,superscriptaddress,longbibliography]{revtex4-1}
\usepackage[english]{babel}
\usepackage{amssymb,amsfonts,amsmath}
\usepackage{array}
\usepackage{dcolumn}
\usepackage{longtable}
\usepackage{hyperref}
\usepackage{float}
\usepackage{gensymb}
\usepackage{bbm}
\usepackage{bm}
\usepackage{graphicx}
\usepackage{datetime}
\usepackage{pifont}
\usepackage{comment}
\usepackage{dsfont}
\usepackage{framed}

\begin{document}
\title{Geometrical origins of contractility in disordered actomyosin networks}
\author{Martin Lenz}
\email{martin.lenz@u-psud.fr}
\affiliation{Univ. Paris-Sud; CNRS; LPTMS; UMR 8626, Orsay 91405 France}

\begin{abstract}
Movement within eukaryotic cells largely originates from localized forces exerted by myosin motors on scaffolds of actin filaments. Although individual motors locally exert both contractile and extensile forces, large actomyosin structures at the cellular scale are overwhelmingly contractile, suggesting that the scaffold serves to favor contraction over extension. While this mechanism is well understood in highly organized striated muscle, its origin in disordered networks such as the cell cortex is unknown. Here we develop a mathematical model of the actin scaffold's local two- or three-dimensional mechanics and identify four competing contraction mechanisms. We predict that one mechanism dominates, whereby local deformations of the actin break the balance between contraction and extension. In this mechanism, contractile forces result mostly from motors plucking the filaments transversely rather than buckling them longitudinally. These findings sheds light on recent \emph{in vitro} experiments, and provides a new geometrical understanding of contractility in the myriad of disordered actomyosin systems found \emph{in vivo}.
\end{abstract}

\maketitle

\begin{figure}[b]
\centering\noindent\includegraphics[width=8.7cm]{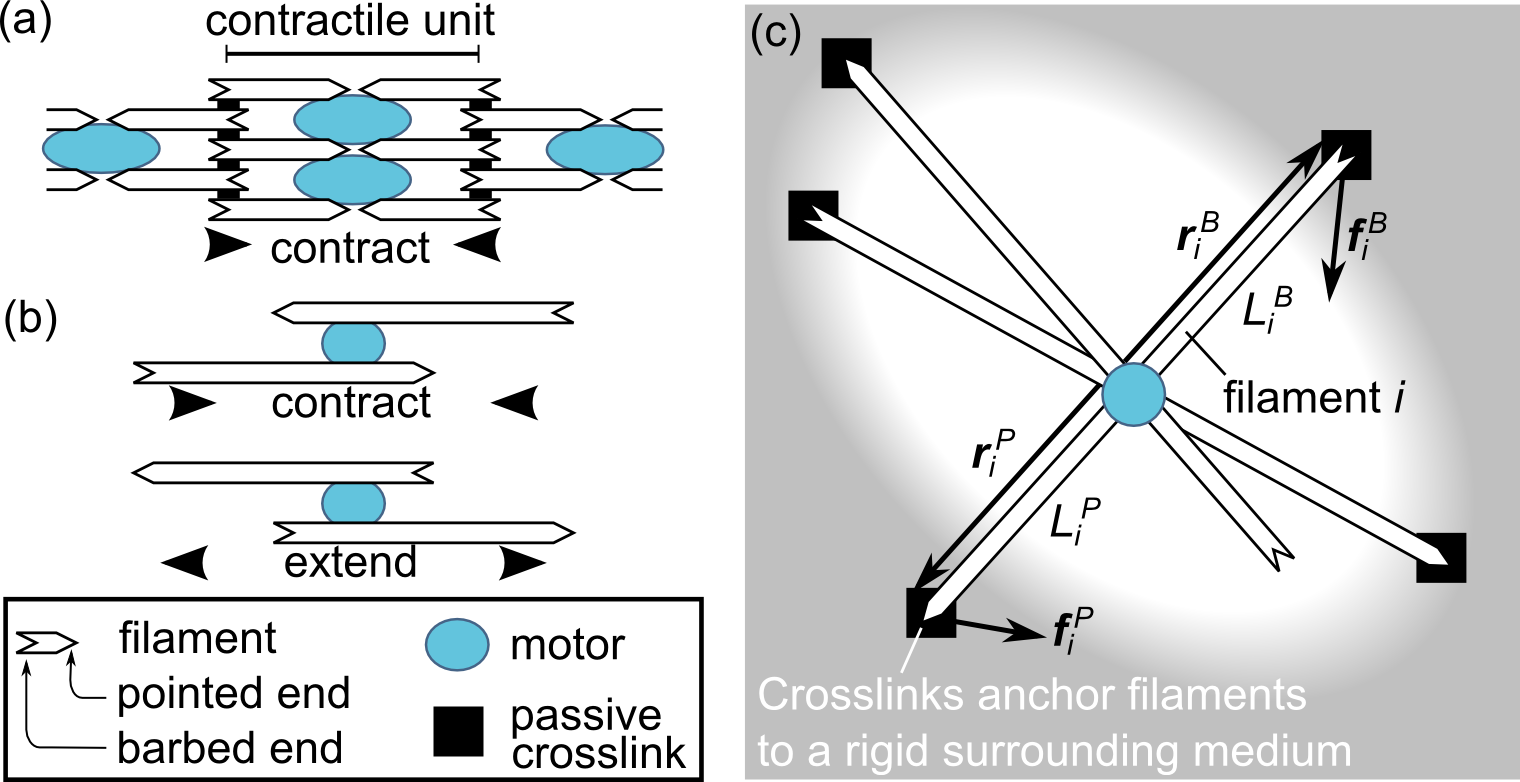}
\caption{\label{fig:nocontraction}Geometrical foundations of contractility. Motors bound to filaments slide towards their ``barbed ends'', as for myosin~II thick filaments. (a)~In striated muscle, motors are localized close to the filaments' pointed ends. When activated, every motor pulls in the neighboring filaments and thus induces local contraction. (b)~If filament polarities are not carefully selected, striated muscle-like locally contractile configurations (\emph{top}) are just as likely as extensile ones (\emph{bottom}), and the overall behavior of the actomyosin assembly is unclear. (c)~The symmetry between contraction and extension subsists in a two- or three dimensional network. In this panel and in the remained of the paper, filament extremities may or may not be crosslinked to the surrounding medium. Even though this is not represented here, crosslinked filaments extend beyond the crosslinks and further into this medium and thus cannot freely rotate around these crosslinks.}
\end{figure}

The structure and motion of living cells is largely controlled by the continuous remodeling of their cytoskeleton, which crucially involves the contractility of networks of actin filaments (F-actin) and myosin molecular motors. How macroscopic motion emerges from the protein-scale interactions between these components was first understood in the context of striated muscle~\cite{Szent-Gyorgyi:2004}. There, individual myosins are assembled into so-called ``thick filaments'', bottlebrush-shaped clusters of myosin capable of binding several actin filaments and of sliding along them for long distances---for brevity we refer to them as ``motors'' in the following. In striated muscle, F-actin and motors are strongly organized into a periodic array of so-called sarcomeres, contractile units where the sliding action of the motors is harnessed to produce contraction through F-actin's geometrical arrangement [Fig.~\ref{fig:nocontraction}(a)].

However, in many biological situations contractile F-actin and myosin assemblies---be they one-dimensional bundles or two- or three-dimensional networks---lack the organization found in sarcomeres~\cite{Fay:1983,Cramer:1997,Carvalho:2009,Medalia:2002,Verkhovsky:1995,Aratyn-Schaus:2011,Salbreux:2012}. While the biochemical processes inducing the relative motion of the motors and filaments are similar to the ones involved in striated muscle, here the geometrical mechanisms used to convert this relative motion into contraction in the absence of organization are less clear. Indeed, the filaments and motors do not have an intrinsic propensity towards contraction, and can \emph{a priori} yield extension just as easily. Figure~\ref{fig:nocontraction}(b) illustrates this property in a simple one-dimensional example. Most theoretical models of disordered actomyosin contractility circumvent this question by assuming from the onset that motors either induce an average contractile stress in the actomyosin medium~\cite{Joanny:2009} or, in more detailed descriptions, that they give rise to localized contractile force dipoles~\cite{MacKintosh:2008aa}. These studies then typically move on to consider the macroscopic consequences of such mesoscopic behaviors. In contrast, in this paper we adopt a different focus and ask how the contractility emerges from the networks' microscopic components in the first place.

This question is most easily discussed in one-dimensional actomyosin assemblies, \emph{i.e.}, actomyosin bundles. There, \emph{in vitro} experiments demonstrate that sarcomere-like organization is not necessary for contraction~\cite{Thoresen:2011}, and thus that the symmetry between contraction and extension illustrated in Fig.~\ref{fig:nocontraction}(b) is spontaneously broken. Because geometry in one dimension is very simple, there are strong geometrical constraints on the type of mechanisms that can lead to such symmetry-breaking~\cite{Lenz:2012a}. Combining these theoretical constraints with further experiments, we have recently shown that F-actin buckling under longitudinal compression enables contraction by favoring local filament collapse in the absence of sarcomere-like organization~\cite{Lenz:2012}.

The situation in two- and three-dimensional actomyosin networks is more complex than that of bundles. There too, contraction arises in random-polarity, disordered \emph{in vitro} networks~\cite{Murrell:2012,Bendix:2008,Silva:2011}. From a theoretical standpoint, however, geometry in two or three dimensions is considerably richer than in one. As a result, several mechanisms can \emph{a priori} give rise to contraction, and symmetry considerations are less easily exploited than in bundles. Accordingly, a range of mechanisms for the emergence of actomyosin contraction have previously been invoked in different levels of detail, ranging from cartoon pictures~\cite{MacKintosh:2008aa,Mizuno:2007aa} to more quantitative numerical~\cite{Dasanayake:2011} and analytical~\cite{Liverpool:2005} approaches. However, there is no consensus regarding their relative roles in either \emph{in vivo} or \emph{in vitro} actomyosin contractility.

Here we present the first comprehensive comparison of contractility-inducing mechanisms in disordered cytoskeletal networks. We first exploit symmetry considerations in two and three dimensions to identify all possible local contraction mechanisms. We then study them individually and compare their relative magnitudes, thus determining the dominant cause of contractility as a function of experimental conditions. Filament deformation is found to play a crucial role in most relevant regimes.

\section{\label{sec:model}Requirements for contraction}
We first show that unlike in striated muscle, filament sliding alone is not sufficient to induce contraction in disordered networks. 
We do this by studying a minimal, sliding-only model and demonstrating that it cannot yield contractility.

We consider a single motor bound to multiple filaments. The filaments are themselves crosslinked to a surrounding rigid external medium as illustrated in Fig.~\ref{fig:nocontraction}(c). We show that overall network contraction cannot occur under the following main assumptions:
\begin{enumerate}\setlength{\itemsep}{0pt}
\item The motor stall force does not depend on its position
\item The motor is point-like
\item The motor is undeformable
\item \label{it:rigid} Filaments behave as rigid rods.
\end{enumerate}
The essence of our argument is as follows. In a network, individual motors may exert either contractile or extensile local forces depending on the polarities of the neighboring filaments [as in Fig.~\ref{fig:nocontraction}(b)]. In a disordered system satisfying the above assumptions, there are as many contractile as extensile motors and the forces produced by the former exactly compensate those produced by the latter. Therefore, the network does not contract \emph{overall}. Thus overall disordered actomyosin contractility requires the breaking of at least one of these assumptions.

We first introduce some notation. The overall contractility of a rigid disordered network is characterized by the average local force dipole~\footnote{More rigorously, $\cal D$ is the trace of a second-rank force dipole tensor $D_{\mu\nu}=\sum_{i,a}(r_i^a)_\mu (f_i^a)_\nu$. Far-field contractility is characterized by the integrated radial stress exerted on a far away sphere. As a rotationally invariant scalar linear in the $f_i^a$s due to the linear elasticity of the rigid medium, this quantity can depend on $D_{\mu\nu}$ only through ${\cal D}=D_{\mu\mu}$.}  $\cal D$ exerted by an individual motor, where
\begin{equation}\label{eq:dipoledef}
{\cal D}= \sum_{i}\sum_{a=B,P}\bm{r}_i^a\cdot\bm{f}_i^a.
\end{equation}
Here $i$ indexes the filaments as in Fig.~\ref{fig:nocontraction}(c), $a=B,P$ denotes the filaments' barbed and pointed ends respectively; therefore each term of the double sum over $i$ and $a$ corresponds to a filament section in contact with the motor. For instance, for the example of Fig.~\ref{fig:nocontraction}(c) $i\in\lbrace 1,2,3\rbrace$, and thus the sum has 6 terms. The position vector of a crosslink is denoted as $\bm{r}_i^a$ and $\bm{f}_i^a$ is the force exerted on it by filament $i$ [Fig.~\ref{fig:nocontraction}(c)]. A negative (positive) $\cal D$ denotes a contractile (extensile) system. The portion of filament between the motor and crosslinker $(i,a)$ is referred to as a ``filament section'' and we denote its length by $L_i^a$.  At steady-state, the motor exerts a longitudinal ``stall force''~$f$ directed towards to pointed end of each filament. This force is transmitted to the crosslinkers through the stretching and compression of the rigid filaments. We thus introduce the stretching moduli $k(L_i^a)$ of the filament sections, \emph{i.e.}, their longitudinal hookean spring constants. In general, $\cal D$ is a function of $f$, the $L_i^a$s and the $k(L_i^a)$s.

We now present our argument in more detail. Consider the filament-motor system of Fig.~\ref{fig:nocontraction}(c). For rigid filaments, linear elasticity applies and the forces $\bm{f}_i^a$ exerted on the crosslinkers are proportional to the motor's stall force. Using Eq.~(\ref{eq:dipoledef}) and noting that the $\bm{r}_i^a$ are constants due to the rigidity of the external medium, this implies
\begin{equation}\label{eq:lineardipole}
{\cal D}\propto f.
\end{equation}
Now consider a new system obtained by reversing the filament polarities of the original system---\emph{i.e.}, exchanging the barbed and pointed ends in Fig.~\ref{fig:nocontraction}(c). As polarities are reversed, the motor reverses its sliding direction on each filament, which is equivalent to changing the sign of its stall force: $f^\textrm{reversed}= -f$. Using Eq.~(\ref{eq:lineardipole}), the polarity-reversed force dipole thus is ${\cal D}^\textrm{reversed}=-{\cal D}$. Hence if the original system generates contractile forces, then the polarity-reversed system generates the same amount of extensile forces.

To complete our reasoning, we consider a large-scale disordered network comprising many filament-motor systems embedded in a rigid medium. The rigid medium can be described as linearly elastic, and thus the network's overall contractile dipole is proportional to the average dipole of a filament-motor system. Due to the network's disorder, any individual filament-motor system is just as likely to occur as its polarity-reversed counterpart. Averaging the force dipoles over the whole network, we thus find that individual contractile and extensile dipoles cancel mutually. From this we conclude that the network has an overall vanishing contractile force dipole, which completes our proof.

This result is quite general, as it requires only a minimal form of disorder, namely polarity-reversal symmetry (\emph{i.e.}, the property that any arrangement of filaments is just as likely as its polarity-reversed counterpart). This is a variant of a more powerful argument valid for one-dimensional bundles~\cite{Lenz:2012a}; a more formal presentation is given in the Supporting Information. Interestingly, this polarity-reversal symmetry can be broken not only through sarcomeric organization, which yields contractility, but also in solution through a dynamical process of motor-filament coalescence and sliding, which favors extension~\cite{Sanchez:2012}. However this process is not relevant for the rigid networks considered here.

\section{\label{sec:forcedipoles}Competing contractility mechanisms}
While the model considered in the previous section cannot generate contractility, such contractility is experimentally observed in actomyosin networks~\cite{Thoresen:2011,Murrell:2012,Silva:2011,Reymann:2012}. This discrepancy implies that this model is an oversimplification: one or several of its assumptions must be violated. By successively relaxing each of these assumptions, here we systematically review all essential contraction mechanisms and predict the magnitude of the associated contractile forces.

\subsection{\label{sec:barbed}Position-dependent stall force} 
Early models of non-sarcomeric actomyosin bundles~\cite{Kruse:2000,Kruse:2003aa} and networks~\cite{Liverpool:2005} proposed that motors stop upon reaching the filament barbed ends, staying there for some time before eventually detaching. Although experimental evidence for this behavior in actomyosin is lacking, the resulting accumulation of immobile motors at the filament barbed ends would generate sarcomere-like crosslinking [Fig.~\ref{fig:nocontraction}(a)] and thus favor contraction.

\begin{figure}[t]
\centering\noindent\includegraphics[width=69mm]{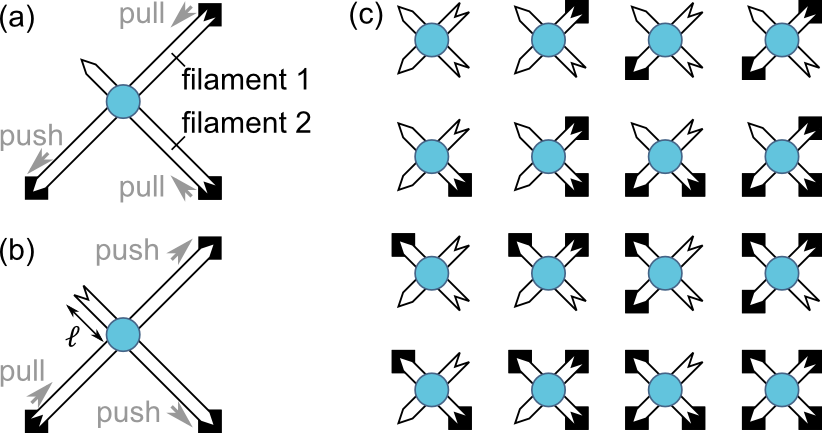}
\caption{\label{fig:dwelling}Contraction induced by a position-dependent stall force. As in Fig.~\ref{fig:nocontraction} and in the following, \emph{black squares} and \emph{blue circles} represent crosslinks and motors, respectively. (a)~Motors in the vicinity of a pointed end typically induce an overall contractile (pulling) force dipole as indicated by \emph{grey arrows} representing the projection of the forces on the direction of the filaments. (b)~Motors close to a barbed end have the opposite effect. (c)~We characterize the resulting net contractility by averaging over all possible local crosslinking configurations.}
\end{figure}

We consider a two-filament system where the motor operation has such a dependence on its distance $\ell$ from the barbed end [Fig.~\ref{fig:dwelling}(a-b)]. Specifically, we assume that the stall force exerted on a filament vanishes~\footnote{To understand this decrease in the stall force, consider that the motor has a position-dependent force-velocity relationship. At any given position, motor velocity decreases with increasing opposing force. In the middle of the filament, the velocity in the absence of force is substantial, and correspondingly the stall force is significantly larger than zero. As the barbed end of the filament is approached, the velocity at zero force vanishes, implying that the motor's stall force goes to zero.} as the motor approaches its barbed end closer than a distance $d\ll\xi$:
\begin{equation}\label{eq:motorweakening}
f(\ell)=f\left(1-e^{-\ell/d}\right).
\end{equation}
The force dipole exerted by a specific configuration depends on whether each of its filament ends is crosslinked to the surrounding medium. For instance, we compute the force dipole associated with Fig.~\ref{fig:dwelling}(a) by resolving force balance under the assumption that the passive crosslinks impose clamped boundary conditions:
\begin{equation}\label{eq:oneconfigdipole}
{\cal D}=-f(\ell_2)L_2^B-f(\ell_1)\frac{L_1^{B}{k}(L_1^{B})-L_1^{P}{k}(L_1^{P})}{{k}(L_1^{P})+{k}(L_1^{B})},
\end{equation}
where $\ell_1$ and $\ell_2$ are the distances from the motor to the barbed ends of filament~1 and 2, respectively. The first term of the right-hand side of Eq.~(\ref{eq:oneconfigdipole}) is always negative, indicating that filament~2 transmits the stall force $f(\ell_2)$ to the bottom-right crosslink, exerting only pulling forces. In contrast, the second term can be either positive or negative as filament~1 distributes this force across two crosslinks and thus exerts both pulling and pushing force. Note that Eq.~(\ref{eq:oneconfigdipole}) is derived in the rigid filament limit $\epsilon=f/\xi k(\xi)\rightarrow 0$, where $\xi$ is the average distance between motor and neighboring crosslinker.

Similar to our derivation of Eq.~(\ref{eq:oneconfigdipole}), we compute the expressions of the force dipoles associated with each possible motor-crosslinker configuration [Fig.~\ref{fig:dwelling}(c)].
Assuming that both the motor and the crosslinkers are uniformly distributed on the filaments, we use these expressions to compute the force dipole averaged over all possible configurations and over filament section lengths:
\begin{equation}\label{eq:dwelldipole}
\langle{\cal D}_\textrm{dwell}\rangle\underset{d\ll\xi\ll{L_f}}{\sim}-\frac{2d}{L_f}f\xi,
\end{equation}
where $L_f$ is the total length of a filament. The condition $L_f\gg\xi$ guarantees that filaments are crosslinked several times and therefore not free to rotate.

To understand why the dipole of Eq.~(\ref{eq:dwelldipole}) is contractile, we remind ourselves that if the stall force were the same irrespective of motor position, the contractile force dipole of Fig.~\ref{fig:dwelling}(a) would exactly cancel the extensile dipole of its polarity-reversed image Fig.~\ref{fig:dwelling}(b). According to Eq.~(\ref{eq:motorweakening}), however, the motor in Fig.~\ref{fig:dwelling}(b) exerts a weaker force on filament~2 than in Fig.~\ref{fig:dwelling}(a) due to the proximity of the filament barbed end. The contractility of Fig.~\ref{fig:dwelling}(a) thus exceeds the extensility of Fig.~\ref{fig:dwelling}(b), resulting in overall contractility. The corresponding average force dipole Eq.~(\ref{eq:dwelldipole}) is thus proportional to the probability $d/L_f$ for the motor to be within a distance $d$ of a barbed end, multiplied by the typical force dipole $f\xi$.

\subsection{\label{sec:Carlsson}Finite motor size} 
\begin{figure}[t]
\centering\noindent\includegraphics[width=89mm]{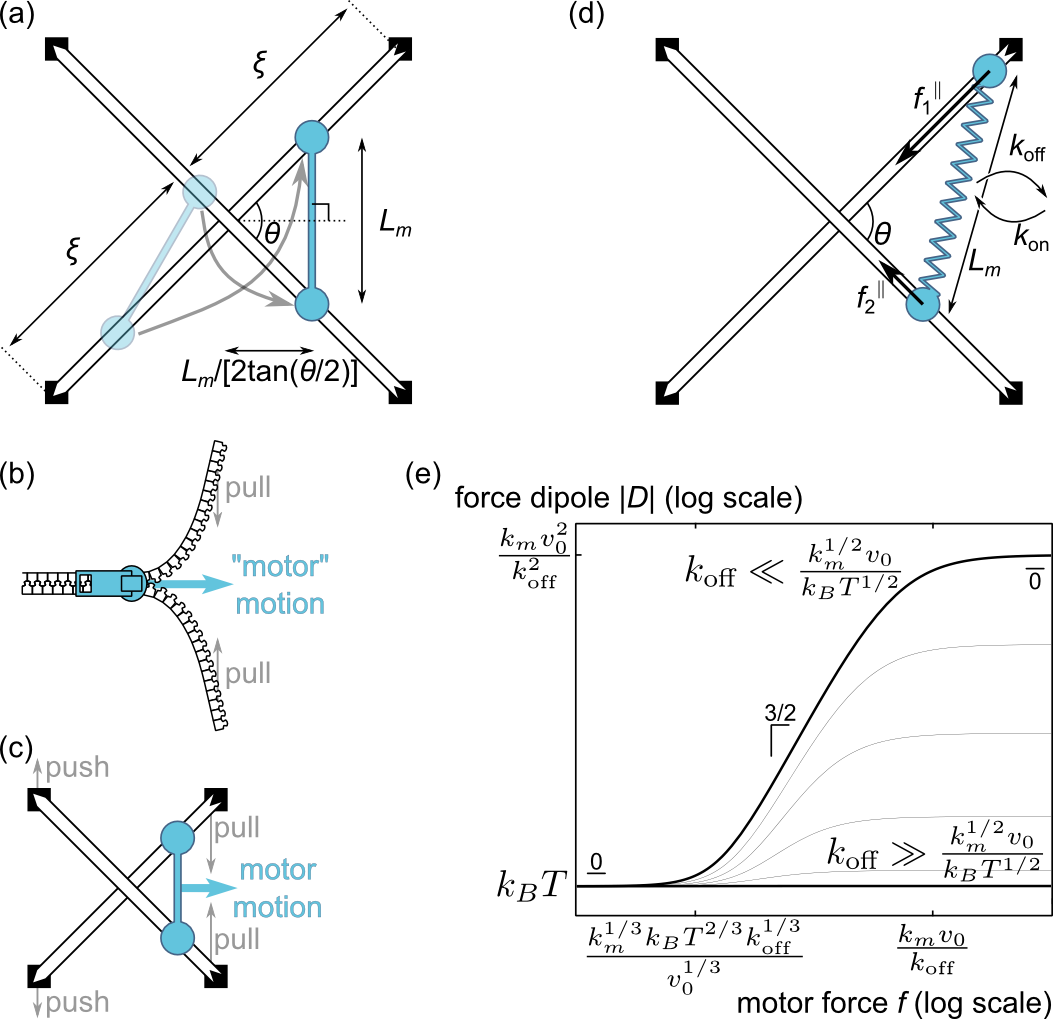}
\caption{\label{fig:finitesize}Contraction induced by finite-size and deformable motors (a)~A finite-size motor minimizes the pseudo-energy Eq.~(\ref{eq:pseudoenergy}) by orienting itself perpendicular to the bisector of the filaments (\emph{dotted line}) as shown by the \emph{grey arrows}. (b)~The contractility induced by such a motor is analogous to the closing force (\emph{thin gray arrows}) of a zipper when its slider is being slid shut (\emph{thick cyan arrow}). (c)~In practice, the zipper-like pulling forces exerted at the barbed end crosslinks are partially compensated by pointed end pushing forces. (d)~An attaching-detaching flexible motor generates contractility in a similar fashion. (e)~Scaling regimes for the deformable motor dipole Eq.~(\ref{eq:Silkedipole}). \emph{Black lines} present the limits of small (\emph{top curve}) and large (\emph{bottom curve}) detachment rate $k_\textrm{off}$ and \emph{thin grey lines} display intermediate regimes.}
\end{figure}

Unlike the point-like motors considered above, a finite-size motor bound to two filaments is not constrained to remain at their intersection. It tends to move towards their barbed ends as shown in Fig.~\ref{fig:finitesize}(a). This motion breaks the equivalence between barbed and pointed end ({aka} polarity-reversal symmetry), thus enabling contraction~\cite{Dasanayake:2011}.

We consider two filaments intersecting at an angle $\theta$ as in Fig.~\ref{fig:finitesize}(a). All filament sections are crosslinked, have length $\xi$ and are considered rigid. The motor is modeled as a rigid dumbbell of length $L_m$ whose heads slide on the filaments until their stall force is reached. To enforce this condition, we minimize the pseudo-energy~\cite{Dasanayake:2011}
\begin{equation}\label{eq:pseudoenergy}
E_m=-f\left(L_1^P+L_2^P\right)
\end{equation}
under the constraint of constant $L_m$. Once the motor is stalled, the mid-point of the motor is offset from the filament intersection by a distance $L_m/[2\tan(\theta/2)]$. Computing the force dipole ${\cal D}(\theta)$ from force balance as in the previous section, we find that small values of $\theta$ yield large motor displacements and thus large force dipoles. We average this force dipole over angles in three dimensions using ${k}(L)\propto L^{-4}$, as expected for filaments with predominantly entropic elasticity~\cite{Odijk:1995}\footnote{The assumption of entropic elasticity is justified for filaments subjected to forces much smaller than their buckling force, which is implicit in the rigid filament assumption used here.}:
\begin{equation}\label{eq:avgfinitesizedipole}
\langle{\cal D_\textrm{finite size}}\rangle= \frac{1}{2}\int_0^{\pi}{\cal D_\textrm{finite size}}(\theta)\sin\theta\,\textrm{d}\theta\underset{L_m\ll\xi}{\sim}-16fL_m.
\end{equation}

To understand the source of this contractile dipole, we draw an analogy between the motor and the slider of a zipper [Fig.~\ref{fig:finitesize}(b-c)]. Assimilating the motor's propensity to slide along the filaments to a closing force applied on the zipper tab, we see that the motor pulls the filament barbed ends together as it progresses, just like the two sides of the zipper chain are pulled together as the zipper closes. This induces a predominantly contractile force dipole.

Importantly, this zipper effect induces contraction only if the motor is displaced from the intersection of the filaments as is the case for a finite-size motor. Indeed, while the motor pulls on the filaments' barbed end crosslinks, it also pushes out on the pointed end crosslinks as shown on Fig.~\ref{fig:finitesize}(c). These two effects compensate exactly for vanishing motor length $L_m=0$, suggesting that for small $L_m$ $\cal D$ is generically proportional to $L_m$. Additionally, $\cal D$ is proportional to $f$ in the rigid filament limit as discussed above. We thus expect zipper-like contractility to scale as
\begin{equation}\label{eq:zipper}
{\cal D}\approx -fL_m,
\end{equation}
consistent with the result of Eq.~(\ref{eq:avgfinitesizedipole}).

\subsection{\label{sec:Silke}Deformable motor} 
We now consider a variant of the previous model where an initially point-like motor can be stretched to a non-zero size, again implying zipper-like contractility. We also consider motor attachment and detachment, as experiments indicate that it can have a significant influence on force build-up in the regimes where the present mechanism will eventually be found to dominate~\cite{Lenz:2012}.

We consider the geometry of Fig.~\ref{fig:finitesize}(d) with a motor of variable length $L_m$ and an associated stretching energy $E_s={k_m}L_m^2/2$, where $k_m$ plays the role of a motor ``spring constant''. The motor detaches from the filaments at a fixed rate $k_\textrm{off}$ and reattaches with $k_\textrm{on}=k_\textrm{on}^0\exp(-E_s/k_BT)$, thus satisfying detailed balance. This rate is substantial only in the region where $E_s \approx k_BT$, implying a motor length $L_m\approx\sqrt{k_BT/k_m}$ of the order of a detached motor's root-mean-square thermal extension. We define the ratio $\eta={\sqrt{k_BT/k_m}/\xi}$ of typical motor size to filament section length and consider the stiff motor limit $\eta\ll1$, analogous to the $L_m/\xi\ll 1$ regime considered above. The velocity $v_i$ of motor head $i$ depends on the projection $f_i^\parallel$ of the motor tension onto the direction of the filament through its force-velocity relationship, assumed linear for simplicity:
\begin{equation}\label{eq:forcevelocity}
v_i=v_0\left(1-f_i^\parallel/f\right),
\end{equation}
where $v_0$ is the motor's unloaded velocity. Taking into account the stochastic attachment/detachment of the motor and its sliding under thermal agitation, we calculate the probability to find it in a given position on the filaments and average the resulting steady-state force dipole over all angles $\theta$ in three dimensions (see Supporting Information). We find
\begin{equation}\label{eq:Silkedipole}
\langle{\cal D}_\textrm{ext}\rangle=-8\pi k_BT\left[
1+{\beta^2}\frac{\sqrt{2+\alpha}-\sqrt{1+\alpha}}{\sqrt{\alpha(1+\alpha)(2+\alpha)}}
\right],
\end{equation}
where $\alpha=k_\textrm{off}f/{2v_0k_m}$ is the ratio of the time required to reach stall to the spontaneous detachment time and $\beta={f}/{\sqrt{k_mk_BT}}$ is the ratio of the motor stall force to the force scale over which the attachment rate varies. The two terms in the square brackets of Eq.~(\ref{eq:Silkedipole}) correspond to two different origins for contractility. We denote the first, $\beta$-independent term as ${\cal D}_\textrm{ext}^\textrm{passive}$. This term does not involve the motor stall force and describes the equilibrium effects of motor binding, which tends to pull the filaments together and exert a contractile force dipole
\begin{equation}\label{eq:Dpassive}
{\cal D}_\textrm{ext}^\textrm{passive}\approx -k_BT.
\end{equation}
The second term, denoted here by ${\cal D}_\textrm{ext}^\textrm{active}$, has two distinct asymptotic regimes. If $\alpha\gg 1$, the motor spontaneously detaches long before reaching stall, yielding a typical extension $L_m\approx v_0/k_\textrm{off}$. In this regime, the motor exerts a typical force $\approx k_mL_m$ on the filaments, equal to the tension of the spring. The resulting typical force dipole is given by the relationship Eq.~(\ref{eq:zipper}) as
\begin{equation}\label{eq:Dactivelargealpha}
{\cal D}_\textrm{ext}^\textrm{active}\underset{\alpha\gg 1}\approx -(k_mL_m)\times L_m\approx -k_mv_0^2/k_\textrm{off}^2.
\end{equation}
Conversely, if $\alpha\ll 1$ the motor reaches stall for moderate angles, implying a force $f$ and an extension $L_m=f/k_m$. However, in this case the average force dipole is not dominated by moderate angles, but rather by small angle configurations for which $\theta\approx\sqrt{\alpha}$. In these configurations, the two filaments are so close to parallel that the motor can slide without stalling until its spontaneous detachment. Similar to the typical motor of the $\alpha\gg 1$ regime, these motors have $L_m\approx v_0\theta/k_\textrm{off}$ and a spring force $\approx k_mL_m$. In the regime $\theta\approx \sqrt{\alpha}$, this yields a force dipole ${\cal D}_\textrm{ext}^\textrm{active}(\theta\approx\sqrt{\alpha})\approx -k_mv_0^2\alpha/k_\textrm{off}^2$. Taking into account the effects of the enlarged recruitment region for small angles [Fig.~\ref{fig:finitesize}(d)], motors in the small-angle regime represent a fraction $\sqrt{\alpha}$ of the total motor population. This leads to an average force dipole
\begin{equation}\label{eq:Dactivesmallalpha}
{\cal D}_\textrm{ext}^\textrm{active}\underset{\alpha\ll1}\approx \sqrt{\alpha}\times{\cal D}_\textrm{ext}^\textrm{active}(\theta\approx\sqrt{\alpha})
\approx
-\frac{f^{3/2}v_0^{1/2}}{k_m^{1/2}k_\textrm{off}^{1/2}}.
\end{equation}
As in the previous section, configurations where the filaments are nearly parallel exert disproportionately large force dipoles which dominate the average.

Figure~\ref{fig:finitesize}(e) ties the asymptotic regimes discussed here together as a function of the original model parameters. In the large detachment rate regime (bottom black curve), detachment is too fast to allow the motors to escape their initial binding region and the force dipole is dominated by its passive component. Conversely, if detachment is slow (top black curve), the magnitude of the motor's stall force matters. The passive dipole still prevails for small forces, while intermediate and large forces are respectively dominated by the active regimes of  Eqs.~(\ref{eq:Dactivesmallalpha}) and (\ref{eq:Dactivelargealpha}).

\subsection{\label{sec:deformation}Deformable filaments}
While the previous sections assumed straight, stiff filaments, here we consider the effect of filament deformation on contractility. Related mechanisms were previously discussed for actomyosin bundles~\cite{Lenz:2012a,Lenz:2012,Liverpool:2009} and gels~\cite{MacKintosh:2008aa,Mizuno:2007aa}. We discuss two asymptotic regimes: small motor forces, which mostly induce filament bending, and large motor forces, which significantly stretch out the filaments' thermal fluctuations. The typical force separating the two regimes is $f\approx k_BT\ell_p^{1/2}/\xi^{3/2}$, \emph{i.e.}, the transverse force required to pull out a significant fraction of these fluctuations.

\subsubsection{\label{sec:smallforce}Small-force regime $f\ll k_BT\ell_p^{1/2}/\xi^{3/2}$} 
\begin{figure}[t]
\centering\noindent\includegraphics[width=88mm]{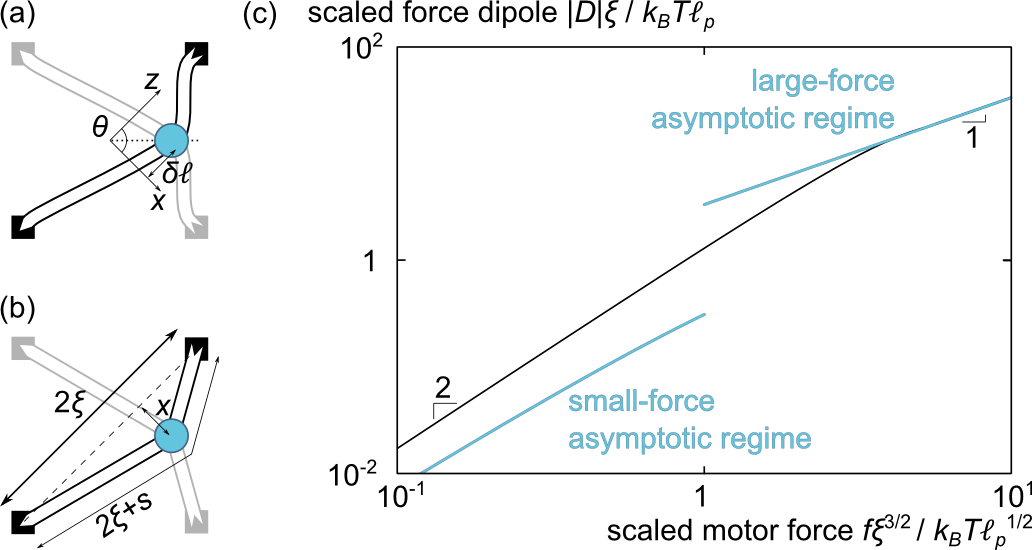}
\caption{\label{fig:bendandstretch}Contraction induced by filament deformation (a)~For small motor forces, the cost of filament deformation is mainly due to bending. The $(x,z)$ coordinate system is given for the darker filament. (b)~For large motor forces, filaments are fully stretched. (c)~Cross-over of the force dipole $\cal D$ between the asymptotic regimes of Eqs~(\ref{eq:benddipole}) and (\ref{eq:stretchdipole}). The interpolating \emph{black line} is discussed in the Supplementary Information.}
\end{figure}

In the absence of significant filament stretching, we consider the filament profile as a weakly perturbed straight line described by the worm-like chain model [Fig.~\ref{fig:bendandstretch}(a)]:
\begin{equation}\label{eq:bendenergy}
E=2\left[\frac{k_BT\ell_p}{2}\int_{-\xi}^{\xi}\left(\frac{\textrm{d}^2x}{\textrm{d}z^2}\right)^2\,\textrm{d}z-f\delta\ell\right],
\end{equation}
where $z$ is the filament's longitudinal direction, $x$ its transverse displacement, $\delta\ell$ the motor's longitudinal displacement and $\ell_p$ the filament persistence length. The last term of Eq.~(\ref{eq:bendenergy}) represents the motor pseudo-energy as in Eq.~(\ref{eq:pseudoenergy}) and contact of the motor with the filaments imposes $x(\delta\ell)=\delta\ell\tan(\theta/2)$.

In this problem, the motor can only progress towards the barbed ends by deforming the filaments. The amplitude $x$ of this deformation is obtained by balancing the filament and motor forces, implying that the filament and motor (pseudo-)energies are of comparable magnitudes and so that $x\approx f\xi^3/(k_BT\ell_p)$. The dominant source of contractile forces is different from the zipper-like mechanism discussed above. Here, the displacement of the motor plucks the filament like the finger of the musician does the string of a harp; interestingly, this mode of deformation induces much larger contractile force than filament buckling~\cite{MacKintosh:2008aa,Mizuno:2007aa} in the $\xi\ll\ell_p$ limit. A small transverse displacement $\approx x$ induces a longitudinal strain $\gamma\approx (x/\xi)^2$ along the filament, hence a filament tension $T\approx (k_BT\ell_p^2/\xi^4)\gamma$, where $k_BT\ell_p^2/\xi^4$ is the typical entropic stretching modulus of the filament~\cite{Odijk:1995}. The resulting  force dipole scales as ${\cal D}\approx -T\xi\approx-f^2\xi^2/k_BT$. A detailed calculation (see Supporting Information) reveals that small angles again have a disproportionately large contribution to the average force dipole, adding a (weak) logarithmic correction to the predicted scaling:
\begin{equation}\label{eq:benddipole}
\langle{\cal D}_\textrm{bend}\rangle \underset{f\ll k_BT\ell_p^{1/2}/\xi^{3/2},\,\xi\ll\ell_p}\sim -\frac{3}{16} \frac{f^2\xi^2}{{k_BT}}\ln\left(\frac{k_BT\ell_p^{1/2}}{c_\textrm{bend}f\xi^{3/2}}\right),
\end{equation}
where $c_\textrm{bend}\simeq 0.191859$. This expression holds until the thermal fluctuations of the filament, which are responsible for its elongational compliance, are pulled out. This occurs for $\gamma\approx \xi/\ell_p$, implying that the small-force regime discussed here is defined by $f\ll k_BT\ell_p^{1/2}/\xi^{3/2}$ as indicated in Eq.~(\ref{eq:benddipole}).

\subsubsection{\label{sec:largeforce}Large-force regime $f\gg k_BT\ell_p^{1/2}/\xi^{3/2}$} 
Under strong extension, the entropic fluctuations of the semiflexible filaments are entirely pulled out, freeing an excess length $s\approx\xi^2/\ell_p\ll\xi$ as shown in Fig.~\ref{fig:bendandstretch}(b). The filaments are therefore analogous to inextensible strings of fixed arclength $2\xi+s$, implying a transverse displacement $x\approx \sqrt{\xi s}$. Since the stalled motor exerts a transverse force $f$, force balance along the $x$ direction imposes a longitudinal filament tension $T\approx f\xi/x$. The force dipole is thus essentially equal to $T\xi\approx f\sqrt{\xi\ell_p}$, consistent with the result of a detailed calculation (see Supporting Information):
\begin{equation}\label{eq:stretchdipole}
\langle{\cal D}_\textrm{stretch}\rangle\underset{f\gg k_BT\ell_p^2/\xi^3,\,\xi\ll\ell_p}\sim-c_\textrm{stretch} f\sqrt{\xi\ell_p},
\end{equation}
with a numerical prefactor $c_\textrm{stretch}\simeq 1.73463$.

We illustrate the crossover between the small- and large-force regimes in Fig.~\ref{fig:bendandstretch}(c).

\section{\label{sec:comparison}Relative importance of each mechanism}
To determine the dominant contraction mechanism, we compare the force dipoles induced by each mechanism presented above as a function of two experimentally controllable parameters: the number of myosin heads per myosin thick filament $N$~\cite{Thoresen:2013} and the inter-crosslink length $\xi$. We consider actin filaments with length ${L_f}=5\,\mu$m and persistence length $\ell_p= 10\,\mu$m. The myosin thick filaments have length $L_m=Nl_m$ with $l_m=3\,$nm, unloaded velocity $v_0=200\,\textrm{nm}\cdot\textrm{s}^{-1}$ and stall force $f=Nf_0$. Since motor heads spend only a fraction of their time bound to actin, we estimate $f_0=0.1\,$pN on average. We use $k_m=\mu/L_m$ with $\mu=45\,\textrm{nN}$ a typical protein filament rigidity~\cite{Kojima:1994}. Myosin II has a duty ratio $1-p_d\simeq 4\%$ and a characteristic attachment-detachment time of $\tau_d=3\,$ms~\cite{Rosenfeld:2003}, yielding a motor detachment rate $k_\textrm{off}=p_d^{N}/\tau_d$. Finally, we assume that motors slow down when their distance to the barbed end is comparable to their size: $d=L_m$. 

\begin{figure}[t] 
\centering\noindent\includegraphics[width=8.7cm]{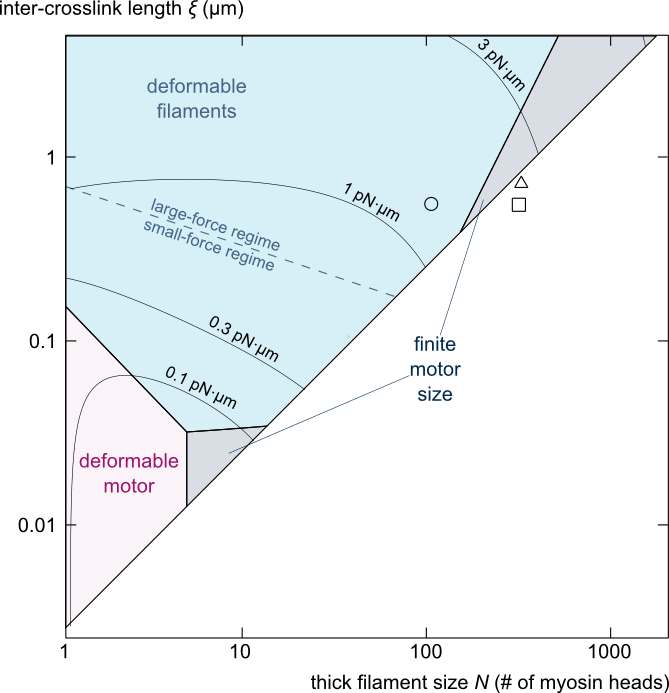}
\caption{\label{fig:phasediag}Contractile forces as a function of experimentally controllable parameters. Colors identify the dominant contraction mechanism in each parameter regime. Contours indicate the magnitude of the contractile force dipole per myosin head $\langle{\cal D}\rangle /N$. 
Symbols indicate the \emph{in vitro} experimental regimes of Ref.~\cite{Silva:2011} (\emph{circle}), Refs.~\cite{Bendix:2008,Koenderink:2009} (\emph{square}) and Ref.~\cite{Alvarado:2013} (\emph{triangle}).
}
\end{figure}

The colored domains in Fig.~\ref{fig:phasediag} indicate as a function of $N$ and $\xi$ which of the four dipoles computed in Sec.~\ref{sec:forcedipoles} has the largest magnitude [Eqs.~(\ref{eq:dwelldipole}), (\ref{eq:avgfinitesizedipole}), (\ref{eq:Silkedipole}) and (\ref{eq:benddipole}-\ref{eq:stretchdipole})]. The bottom-right half of the diagram is left blank as it involves very large motors ($L_m>\xi$) not captured by our current approach; our assumptions $\xi<\ell_p$ and $d<\xi<{L_f}$ are satisfied throughout the top-left (colored) half.  The finite motor size mechanism tends to dominate in the vicinity of the diagonal where the motor size $L_m$ is largest. The deformable motor mechanism dominates in the bottom left corner of the diagram; for these small values of $N$ and $\xi$ and given that the myosin thick filaments are hardly stretchable ($\mu\gg f$), thermal agitation dominates and ${\cal D}_\textrm{ext}^\textrm{active}\ll{\cal D}_\textrm{ext}^\textrm{passive}$. Deformable filament mechanisms govern contractility in large-$\xi$ regions where the filament sections are most flexible and can thus be deformed by motor forces. Finally, the position-dependent stall force mechanism is always negligible in front the finite size motor mechanism; thus it never dominates contractility. This picture is remarkably insensitive on precise parameter values (see Supporting Information).

We next consider the total force dipole $\langle {\cal D}\rangle$, defined as the sum of the four force dipoles computed in Sec.~\ref{sec:forcedipoles}. The magnitude of the total dipole per myosin head $\langle {\cal D}\rangle/{N}$ is represented by contour lines in Fig.~\ref{fig:phasediag}. In the $\xi\gtrsim 0.3\,\mu$m region, these forces compare with the force dipole exerted by a myosin head in striated muscle ${\cal D}/N=(500\,\textrm{pN}\times3\,\mu\textrm{m})/300=5\,\textrm{pN}\cdot\mu\textrm{m}$; 
filament deformation-based mechanisms dominate most of this parameter region. Conversely, for $\xi\lesssim 0.3\,\mu$m forces are much weaker, and possibly too small for experimental observation. Consistent with this, the typical network parameters used in \emph{in vitro} experimental studies of actomyosin contractility are confined to the strong-contractility region (Fig.~\ref{fig:phasediag}, symbols~\footnote{The symbols' coordinates are computed by assimilating $\xi$ to the networks' entanglement length $l_e=620\,\textrm{nm}/c^{2/5}$, with the actin concentration $c$ in units of mg/mL~\cite{Isambert:1996}.}). Interestingly, these symbols lie between the deformable filaments and the finite motor size contraction domains, suggesting that both mechanisms could play a role in these experiments.

\section{\label{sec:discussion}Discussion}
While the emergence of contractility in strongly organized actomyosin assemblies is well understood, here we consider this process in disordered networks such as those found in non-muscle cells. Among all possible local contraction models, actin filament deformation (bending or stretching) is most prominent in favoring locally contractile motor/filament configurations over locally extensile ones. In this mechanism, filament deformation causes contractility rather than being a mere byproduct of it. Local rearrangements due to the motors' finite size could also play a role in \emph{in vitro} experiments. We formulate quantitative predictions of the forces generated by these mechanisms, yielding insights into the influence of the network's microstructure and enabling experimental verifications.

The predicted importance of filament deformation is consistent with \emph{in vitro} studies where the deformation of a reconstituted actomyosin sheet is found to exactly coincide with the amount of deformation of individual filaments, suggesting that filament deformation indeed causes contraction~\cite{Murrell:2012}. We also account for the observed inhibition of contractility by excessive crosslinking ($\cal D$ vanishes for $\xi\rightarrow 0$)~\cite{Bendix:2008}. Additionally, the fact that almost parallel filaments dominate contractility in most of the mechanisms studied here is in good agreement with simulations suggesting that filament alignment favors contraction~\cite{Dasanayake:2013}. It would be interesting to extend our results to partially bundled networks---which readily form \emph{in vitro}~\cite{Falzone:2012}---knowing that contraction within a bundle also crucially involves filament deformation~\cite{Lenz:2012a,Lenz:2012}. Note however that in the mechanism described here motors pull on the filaments both in the longitudinal and transverse direction, while in bundles only longitudinal forces are significant. Consequently, motors pulling tranverse to a bundle might be much more effective at deforming the actin and thus generating contraction than the ones within, as the latter are deforming the filaments through comparatively ineffective buckling. Finally, we note that  \emph{in vitro} parallel bundles of actin filaments contract considerably less than antiparallel bundles~\cite{Reymann:2012}, in contradiction with a robust prediction of the position-dependent stall force model~\cite{Kruse:2000}; this supports our finding that the position-dependent stall force has little effect on contractility. This conclusion could however be modified in networks of, \emph{e.g.}, kinesin motors and the stiff filaments microtubules.

Although we find that filament deformations dominate many significant regimes of actomyosin contraction, our focus on local actin deformation could still lead to an underestimate of their effect. Indeed, nonlocal deformations of the network over several mesh sizes could be more favorable than local deformations in heavily crosslinked networks or regimes where motors are larger than the inter-crosslink length. Collective effects could also be of importance, as stress propagation through the elastic filament network could lead to cooperativity between distant motors. We note that our weakly deformed networks approach is only relevant for small motor forces or during the very early stages of larger-scale contraction. Further work is required to analyze strongly deformed or dynamically reorganizing networks and the corresponding synergies between several of the mechanisms described here. On such longer time scales, the microscopic interactions between filaments and motors considered here could furthermore shed light onto the self-organization of disordered actomyosin networks into more organized structures~\cite{Aratyn-Schaus:2011}.

Assessments of the experimental relevance of the mechanisms described here will be facilitated by recent developments in \emph{in vitro} assays~\cite{Thoresen:2011,Murrell:2012,Silva:2011,Reymann:2012,Thoresen:2013}. Indeed, these now allow precise tuning of the motor and network characteristics as well as detailed monitoring of the network deformations, from which the magnitude of the local force dipole could be inferred.
How these considerations apply \emph{in vivo} is a fascinating question, which requires further investigations into alternatives to the paradigm of sarcomere-like contraction.

\acknowledgements{I thank M. Gardel, M. Murrell and T. Thoresen for countless inspiring discussions as well as P. Ronceray, A. Roux and C. Sykes for careful reading of the manuscript. Our group belongs to the CNRS consortium CellTiss. This work was supported by grants from Universit\'e Paris-Sud and CNRS, the University of Chicago FACCTS program, Marie Curie Integration Grant PCIG12-GA-2012-334053 and ``Investissements d'Avenir'' LabEx PALM (ANR-10-LABX-0039-PALM).}

\end{document}